\documentclass{article}
\usepackage{graphicx}
\usepackage{natbib}
\usepackage{amsmath}
\usepackage[colorlinks=true,urlcolor=blue,linkcolor=blue,citecolor=blue]{hyperref}
\usepackage{subcaption}

\usepackage[normalem]{ulem}
\usepackage{xcolor}
\usepackage{authblk}

\title{Transdimensional surface wave tomography of the near-surface: Application to DAS data}
\author{Amin Rahimi Dalkhani$^1$, Mus'ab Al Hasani$^1$, Guy Drijkoningen$^1$, \
\and Cornelis Weemstra$^{1,2}$\\
$^1$ Department of Geoscience and Engineering, Delft University of Technology, The Netherlands\\
$^2$ Seismology and Acoustics, Royal Netherlands Meteorological Institute, The Netherlands}

\begin{document}

\maketitle
\begin{abstract}
    
    Distributed Acoustic Sensing (DAS) is a novel technology that allows sampling of the seismic wavefield densely over a broad frequency band. This makes it an ideal tool for surface wave studies.
    In this study, we evaluate the potential of DAS to image the near-surface using synthetic data and active-source field DAS data recorded with straight fibers in Groningen, the Netherlands. First, we recover the laterally varying surface wave phase velocities (i.e., local dispersion curves) from the fundamental-mode surface waves. We utilize the Multi Offset Phase Analysis (MOPA) for the recovery of the laterally varying phase velocities. In this way, we take into account the lateral variability of the subsurface structures. 
    Then, instead of inverting each local dispersion curve independently, we propose to use a novel 2D transdimensional surface wave tomography algorithm to image the subsurface. In this approach, we parameterize the model space using 2D Voronoi cells and invert all the local dispersion curves simultaneously to consider the lateral spatial correlation of the inversion result. Additionally, this approach reduces the solution nonuniqueness of the inversion problem. 
    The proposed methodology successfully recovered the shear-wave velocity of the synthetic data. Application to the field data also confirms the reliability of the proposed algorithm. The recovered 2D shear-wave velocity profile is compared to shear-wave velocity logs obtained at the location of two boreholes, which shows a good agreement.
\end{abstract}

\section{Introduction}

After the first field trials of the Distributed Acoustic Sensing (DAS) \citep{bostick2000field,molenaar2012first,johannessen2012distributed}, the DAS found many applications in different geophysical problems \citep{barone2021tackling}. Due to the dense sampling of the wavefield and the broadband character, DAS is getting popular in surface wave studies both in active \citep[e.g., ][]{qu2023trans,yust2023near} and passive \citep[e.g., ][]{ajo2019distributed,nayak2021measurement} surveys. 

A material's shear-wave velocity directly depends on its shear strength (or stiffness). Consequently, the shear-wave velocity of the near-surface is a valuable parameter in many subsurface engineering applications. And because the velocity of surface waves strongly depends on that shear-wave velocity, near-surface shear-wave velocity models are frequently derived from surface wave measurements \citep{socco2010surface}. Typically, two types of surface waves are recorded: Rayleigh and Love waves \citep{Aki:2002rt}. Surface waves are dispersive when the shear-wave velocity varies as a function of depth, meaning that different frequencies propagate  with different velocities. A surface wave's propagation velocity at an individual frequency is referred to as that frequency's `phase velocity'. The first step in surface wave imaging is therefore usually the retrieval of the frequency-dependent phase velocities. An inversion subsequently results in a model of the shear-wave velocity as a function of depth \citep{Schaefer:2011ve,zhang20201}.

Conventionally, in an active-source surface wave analysis, a dispersion curve is retrieved from each common shot record using Multichannel Analysis of Surface Waves \citep[MASW;][]{park1999multichannel} assuming that the subsurface is a stack of horizontal layers. Then, an inversion algorithm is used to recover a 1D shear-wave velocity profile \citep[e.g., ][]{vantassel2022extracting,qu2023trans}.
However, due to the fine spatial sampling and the high-frequency content of data recorded in an active-source survey, DAS recordings enhance the lateral resolution significantly \citep{barone2021tackling}. Consequently, many authors \citep[e.g., ][]{neducza2007stacking,luo2008generation,vignoli2011statistical,barone2021tackling} suggested recovering local dispersion curves (DCs) to take into account the lateral variation of the subsurface velocity structure. These local DCs can then be used in an inversion algorithm to recover a 2D profile of the subsurface shear-wave velocity \citep[][]{vignoli2016frequency,barone2021tackling}. 

Several approaches have been proposed to account for lateral variations of the subsurface's shear-wave velocity. The most common approach involves the application of a moving window to the recorded shot gathers in the time-space domain \citep[e.g., ][]{bohlen20041,luo2008generation,socco2009laterally,boiero2010retrieving}. The windowed part of the data is then transformed into a spectral domain to estimate the ``local" dispersion curve at the location of the center of that window. Alternatively, 
Multi Offset Phase Analysis \citep[MOPA;][]{vignoli2016frequency} can be used to consider lateral variation\citep{barone2021tackling}. \citet{vignoli2016frequency,barone2021tackling} successfully retrieved fundamental-mode laterally varying dispersion curves using MOPA. Then, they inverted the local dispersion curves at each location independently to recover a 2D shear-wave velocity pseudo-section of the subsurface. It is worth noting that both studies \citep{vignoli2016frequency,barone2021tackling} applied the MOPA algorithm to the Geophone recordings.

In this study, we apply the MOPA algorithm to a 2D active-source DAS recording for the first time. We recover the laterally varying fundamental-mode local phase velocities. Then, we propose to invert all the dispersion curves simultaneously using a nonlinear 2D transdimensional tomographic algorithm \citep{bodin2009seismic}. This transdimensional inversion algorithm is originally developed for seismic travel time tomography. Later, it has been used for many geophysical problems \citep[e.g., ][]{dettmer2012trans,rahimi2021potential,ghalenoei2022trans,yao2023trans}. Here, we develop the algorithm for the purpose of active-source surface waves recorded with DAS to image and recover a 2D shear-wave velocity section of the subsurface with improved lateral correlation. Finally, we validate the inverted 2D shear-wave velocity section by comparing it with the shear-wave velocity logs obtained from two boreholes along the acquisition line. 

\section{Theory \& methodology}

In this section, we detail the various methods used in this study. First, we describe how the surface waves are obtained from the fiber optic recordings. Subsequently, we explain how we locally estimate the surface waves' phase velocity as a function of frequency. Finally, we provide the details of the McMC implementation.

\subsection{Distributed acoustic sensing (DAS)}

In this part, we describe what is measured by DAS and how it relates to geophone measurements. Unlike geophones which are point sensors that record the particle velocity at a specific location, DAS measurements are average strain-rate, i.e. $\dot{\varepsilon}_{xx} = \partial_t \left(\partial_x u \right) $, along a specific distance called the gauge length ($L_g$). Alternatively, DAS response can be equivalently represented by the spatial derivative of the velocity vector, i.e. $ \partial_x \left(\partial_t u \right) = \partial_x V_x$ \citep{Daley2016}. 

\begin{figure}
    \centering
    \includegraphics[width=\textwidth]{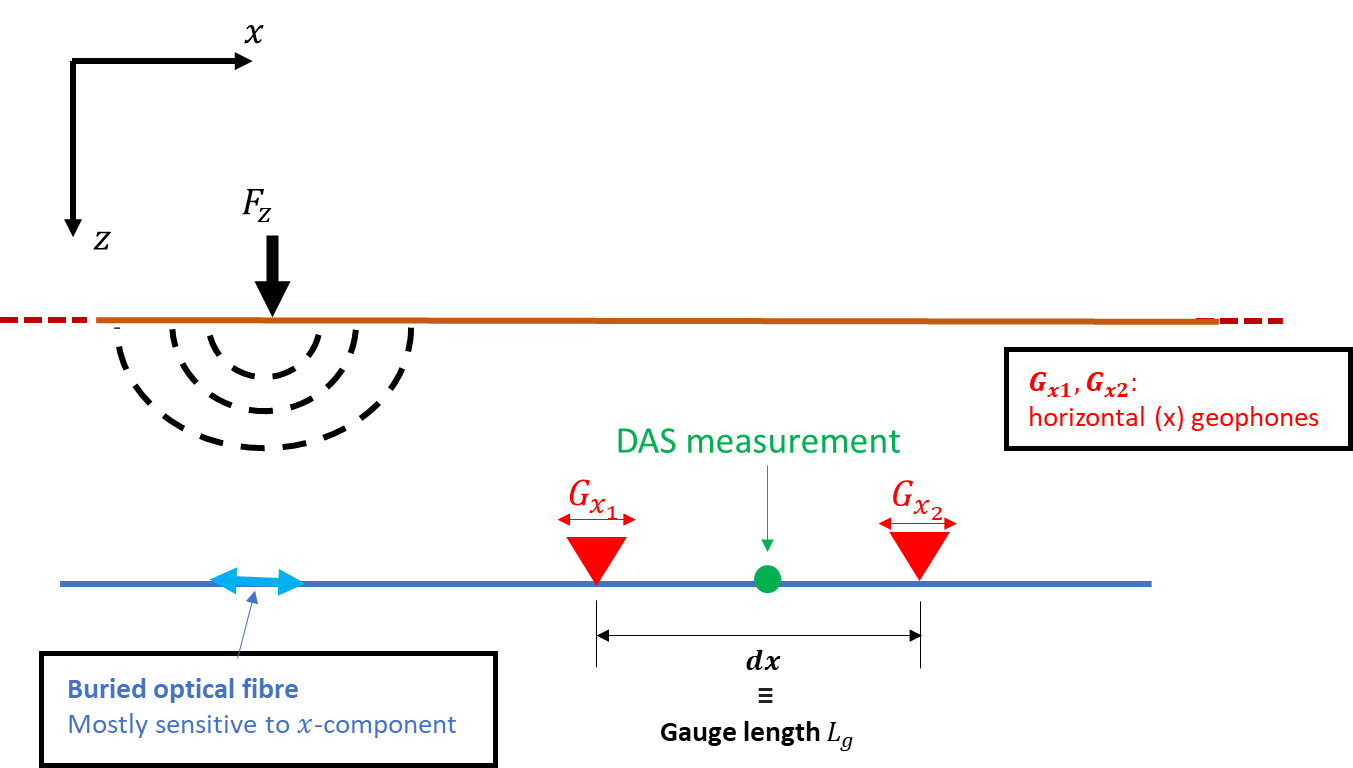}
    \caption{An illustration to compare DAS and geophone measurement.}
    \label{fig:DAS_vs_geoph}
\end{figure}

We show a schematic illustrating a scenario mimicking our implementation in the field experiment in Figure~\ref{fig:DAS_vs_geoph}.  A vertical source $F_z$ will generate different types of waves including surface waves that are mostly Rayleigh waves. As DAS relies on the elongation and contraction of the fibre as a function of time, its maximum sensitivity is along its axial direction. So it is safe to say the majority
of what is being recorded is the horizontal component of the Rayleigh waves. 
To compare what is being measured by geophones to DAS, horizontal ($x$) geophones $G_{x_1}$ and $G_{x_2}$ separated by distance $dx$ is assumed to be recording the horizontal component of the particle velocity $V_x$ at $x_1$ and $x_2$, respectively. Equivalent DAS response between the two geophones can be estimated as the spatial derivative of the particle velocity with the following expression adapted from \cite{Zulic2022}:

\begin{equation}\label{eq:das_resp}
    DAS\left(\frac{x_2 - x_1 }{2}\right) = \frac{V_x(x_2) – V_x(x_1)}{dx},
\end{equation}
where, $dx$ is equivalent to the gauge length $L_g$ of the DAS measurement. Given that the gauge length is small and using dense spatial sampling that to satisfy the spatial Nyquist criterion, the phase velocity is not affected by the use of strain rate instead of the commonly used particle velocity.

\subsection{Phase velocity retrieval}

A variety of methods exist to retrieve a surface wave mode's dispersion curve. The dispersion curve describes the mode's phase velocity as a function of frequency. Often, dispersion curves are estimated by transforming the data into a different (spectral) domain. For example, the recorded data can be transformed from the time-space domain to the phase velocity-frequency \citep[$c-f$; ][]{mcmechan1981analysis} domain, to the frequency-wavenumber \citep[$f-k$; ][]{foti2000notes} domain, or to the phase-offset \citep[$\phi-x$; ][]{strobbia2006multi} domain. These methods have in common that they rely on the assumption that the subsurface is a stack of horizontal layers. In other words, the subsurface is assumed to be laterally invariant. Consequently, in the case of laterally rapidly varying structure, the retrieved dispersion relation is effectively a (non-exact) average \citep{boiero2010retrieving}. In practice, these methods are assumed still to be valid when the subsurface is laterally very smooth.

In case the subsurface is laterally invariant, the relation between the phase ($\phi$) and offset ($x$) will be linear at each discrete frequency ($f_i$), with the slope coinciding with the wavenumber ($k$). This can be formulated as \citep{strobbia2006multi}:
\begin{equation}
    \phi(f_i,x) = k(f_i)x + \phi_0(f_i),
\label{eq:linearphaseoffset}
\end{equation}
where $\phi_0$ is the phase at the location of the source. Equation \eqref{eq:linearphaseoffset} allows the estimation of a wavenumber by means of a least-squares fit of the phase-offset data at each discrete frequency $f_i$, i.e., using a linear regression \citep[for detailed formulation see ][]{strobbia2006multi}. The estimated wavenumber ($k(f_i)$) can be translated to the frequency-dependent phase velocity ($c(f_i)$) using:
\begin{equation}
     c(f_i) =\dfrac{2\pi f_i}{k(f_i)},
     \label{eq:phasevelocitywavenumber}
 \end{equation}
where $i = 1, 2, ..., N_f$, and $N_f$ denotes the number of discrete frequencies. This approach is referred to as the `Multi Offset Phase Analysis' \citep[MOPA; ][]{strobbia2006multi,vignoli2010identification}. 

In the above description, the subsurface is approximated by a single dispersion curve for the whole of $x$. In the presence of lateral variations, Equation~\eqref{eq:linearphaseoffset} can be formulated such that the wavenumber $k$ varies as a function of both offset and frequency. That is, $k=k(f_i,x^{\mathrm{(c)}})$, where $x^{\mathrm{(c)}}$ denotes the center position of a set of adjacent (Fourier transformed) wavefield recordings running from $x^{\mathrm{(c)}}-W/2$ to $x^{\mathrm{(c)}}+W/2$ (here, $W$ is the spatial window along which $k(f_i,x^{\mathrm{(c)}})$ is assumed to be constant). \citet{vignoli2011statistical} ``move" this spatial window along the recording line with small steps (i.e., significantly smaller than $W$). This is done separately for each discrete frequency, allowing $W$ to be frequency dependent \citep{vignoli2016frequency}. The wavenumbers are subsequently derived using linear regression according to Equation \eqref{eq:linearphaseoffset}. These laterally varying wavenumbers are subsequently converted to laterally varying phase velocities $c(f_i,x^{\mathrm{(c)}})$ using Equation~\ref{eq:phasevelocitywavenumber}. Since the lateral resolution of the surface waves is directly related to the wavelength \citep{barone2021tackling}, \citet{vignoli2016frequency} proposed to have the spatial window length $W$ be a function of the wavelength of the surface waves.

In this study, we use the MOPA algorithm of \citet{vignoli2016frequency} to estimate local dispersion curves. Chiefly, this is because of the algorithm's robustness and simplicity. A drawback of the MOPA algorithm, however, is that it can only be applied to a single surface wave mode. Therefore, only one mode will be considered, and the rest of the available surface wave modes and body waves have to be filtered out. Finally, it should be understood that the recovered wavenumbers are associated with the wavefield. That is, they will deviate from the medium's true wavenumber distribution \citep[sometimes referred to as `structural wavenumbers';][]{Wielandt1993}, with the discrepancy between the two being larger for more heterogeneous subsurfaces. 

\subsection{Transdimensional surface wave tomography}

A single dispersion curve, i.e., $c(f_i,x^{\mathrm{(c)}})$ for a specific location $x^{\mathrm{(c)}}$ (with $i = 1, 2, ..., N_f$), can be ``inverted" to recover a (1D) shear-wave velocity profile. 
Numerous inversion algorithms are described in the literature. Roughly speaking, one can distinguish between linearized algorithms \citep[e.g., ][]{xia1999estimation} and nonlinear global search methods. The latter include genetic algorithms \citep{yamanaka1996application}, simulated annealing \citep{beaty2002simulated}, the neighborhood algorithm \citep{wathelet2008improved}, Monte Carlo methdos \citep{socco2008improved}, particle swarm optimization \citep{wilken2012application}, and a 1D transdimensional algorithm \citep{bodin2012transdimensional}.

In the presence of lateral variations, the dispersion relation varies as a function of location. In that case, the different dispersion curves are often inverted independently using one of the mentioned 1D inversion algorithms \citep[e.g., ][]{bohlen20041,socco2009laterally,vignoli2016frequency,barone2021tackling}, after which the independently inverted 1D profiles are pieced together to obtain a 2D (or 3D) shear-wave velocity pseudo-section (or pseudo-cube). However, by independently inverting the adjacent dispersion curves, lateral correlations in the subsurface structure are ignored \citep{zhang20201}. \citet{socco2009laterally} propose to invert all dispersion curves simultaneously to mitigate the solution's non-uniqueness, and retain lateral smoothness. They use a laterally constrained least-squares algorithm in which each 1D model is linked to its neighbors. \citet{zhang20201} invert all dispersion curves simultaneously using a (Bayesian) 3D transdimensional algorithm. In this study, we invert all $c(f_i,x^{\mathrm{(c)}})$ simultaneously using a 2D transdimensional algorithm \citep{bodin2009seismic}. As such, we retain lateral shear-wave correlations and circumvent (rather arbitrary) smoothing and damping procedures.

The 2D transdimensional tomographic algorithm by \citet{bodin2009seismic} is originally developed for travel time tomography. Here, we modify the algorithm to invert all DAS-derived dispersion curves simultaneously. Our transdimensional algorithm uses a 2D Voronoi tessellation to parameterize the model space ($\mathbf{m}$) in combination with a reversible jump Markov chain. A Voronoi cell is defined by the location of its nucleus and the shear-wave velocity assigned to it. The geometry of each cell is controlled by its neighboring cells. Reversible jump Markov chain Monte Carlo \citep[rjMcMC; ][]{green1995reversible} allows a variable parameterization of the model space, meaning that the number of Voronoi cells, their locations, and the assigned velocities are all unknowns \citep{rahimi2021potential}. The transdimensional parameterization allows the algorithm to sample the full model space, without the need to introduce any kind of regularization \citep{bodin2009seismic}. 

The rjMcMC algorithm is a Bayesian inference method that aims to sample the posterior probability density of the model parameters given the observed data, $p(\mathbf{m}  | \mathbf{d} )$. The posterior is proportional to the product of the likelihood $p(\mathbf{d} | \mathbf{m})$ and the prior $p(\mathbf{m})$ \citep{bodin2009seismic, rahimi2021potential}:
\begin{equation}
        p(\mathbf{m}  | \mathbf{d} ) \propto p(\mathbf{d} | \mathbf{m})p(\mathbf{m}).
\end{equation}
The prior probability distribution, $p(\mathbf{m})$, incorporates all (a priori) known independent information about the model space. Similar to \citet{bodin2009seismic}, we consider an uninformative uniform prior for all the model parameters (i.e., number of cells, Voronoi nuclei location, and velocity assigned to each cell). 

The likelihood function $p(\mathbf{d} | \mathbf{m})$ plays a fundamental role in the inference of the model space as it provides the probability of the observed laterally varying dispersion curves given a specific velocity model. It is formulated as: 
\begin{equation}
p(\mathbf{d} | \mathbf{m}) =\prod_{i=1}^{N_f}\prod_{j=1}^{N_x}\bigg(\dfrac{1}{\sqrt{2\pi}\sigma_{ij}} \exp(-\dfrac{\big(g_{ij}(\mathbf{m})-d_{ij}\big)^2}{2\sigma_{ij}^2})\Bigg),
\label{eq:Likelihood}
\end{equation}
where $N_x$ is the number of locations $x_j^{\mathrm{(c)}}$ for which a dispersion curve is estimated (i.e., $j = 1, 2, ..., N_x$). Data point ${d}_{ij}$ is the phase velocity at discrete frequency $f_i$ and location $x_j^{\mathrm{(c)}}$ (Figure~\ref{fig:MOPA_Syntheticdata}e). The vector $\mathbf{m}$ contains the parameters describing the proposed model. Due to the variable number of Voronoi cells, its length changes while the posterior is being sampled. Furthermore, $g_{ij}$ is the modeled laterally varying phase velocity and $\sigma_{ij}$ is the data uncertainty or noise level for the phase velocity at the discrete frequency $f_i$ and location $x_j^{\mathrm{(c)}}$.

The reversible jump Markov chain draws samples from the posterior distribution by means of a Metropolis-Hasting (MH) algorithm which includes changing the dimension of the model space. Jumping between different dimensions of the model space allows the rjMcMC algorithm to perform a global search and overcome the problem of local minima \citep{andrieu1999robust}. The process starts with some random initial model $\mathbf{m}$. Then, the algorithm draws the next sample of the chain by proposing a new model, $\mathbf{m}^\prime$, based on a known proposal probability function, $q(\mathbf{m}^\prime|\mathbf{m})$, which only depends on the previous state of the model $\mathbf{m}$. The proposed model $\mathbf{m}^\prime$ will be accepted with probability \citep{bodin2009seismic}:
\begin{equation}
     \alpha(\mathbf{m^\prime}|\mathbf{m})  =min[1,\dfrac{p(\mathbf{m^\prime})}{p(\mathbf{m})}\dfrac{p( \mathbf{d}|\mathbf{m^\prime})}{p( \mathbf{d}|\mathbf{m})}\dfrac{q(\mathbf{m} | \mathbf{m^\prime})}{q(\mathbf{m^\prime} | \mathbf{m})} \times |\mathbf{J}|],
     \label{eq:acceptanceprobability}
\end{equation}
where, $\dfrac{p(\mathbf{m^\prime})}{p(\mathbf{m})}$ is the prior ratio, $\dfrac{p( \mathbf{d}|\mathbf{m^\prime})}{p( \mathbf{d}|\mathbf{m})}$ is the likelihood ratio, $\dfrac{q(\mathbf{m} | \mathbf{m^\prime})}{q(\mathbf{m^\prime} | \mathbf{m})}$ is the proposal ratio, and $\mathbf{J}$ is the Jacobian of transformation from $\mathbf{m}$ to $\mathbf{m^\prime}$ and is needed to account for scale changes involved when the perturbation considers a jump between dimensions \citep{green1995reversible}. 

The acceptance probability, $\alpha(\mathbf{m}^\prime|\mathbf{m})$, is the key to ensuring that the samples will be generated according to the target posterior distribution, $p(\mathbf{m} | \mathbf{d})$ \citep{bodin2009seismic}. 
Similar to \citet{bodin2009seismic}, we used four perturbation types to propose a new model ($\mathbf{m}^\prime$) based on the current model ($\mathbf{m}$) including nuclei move, velocity update, birth, and death steps. We followed exactly the same way as \citet{bodin2009seismic} to parametrize the model space and perturb the model space using the four perturbation types. Consequently, the formulas to compute the acceptance probability (Equation \ref{eq:acceptanceprobability}) are identical to the ones derived by \citet{bodin2009seismic}. 

It is worth noting that the input data in our case (i.e., laterally varying phase velocities along a 2D line) is different from the travel times used in \citet{bodin2009seismic}. Consequently, a different forward function ($g$ in Equation \ref{eq:Likelihood}) is necessary to compute the modeled data. For this purpose, we use a MATLAB package developed by \citet{wu2019matlab} using the method proposed in \citet{buchen1996free}. This algorithm computes the dispersion curve (i.e., phase velocity versus frequency) in a 1D earth model. Therefore, to model the laterally varying dispersion curves, we take the 1D velocity profile at each location and compute the dispersion curves independently.

\section{Application to synthetic data}

\begin{figure}[h]
    \centering
    \includegraphics[width=1\linewidth]{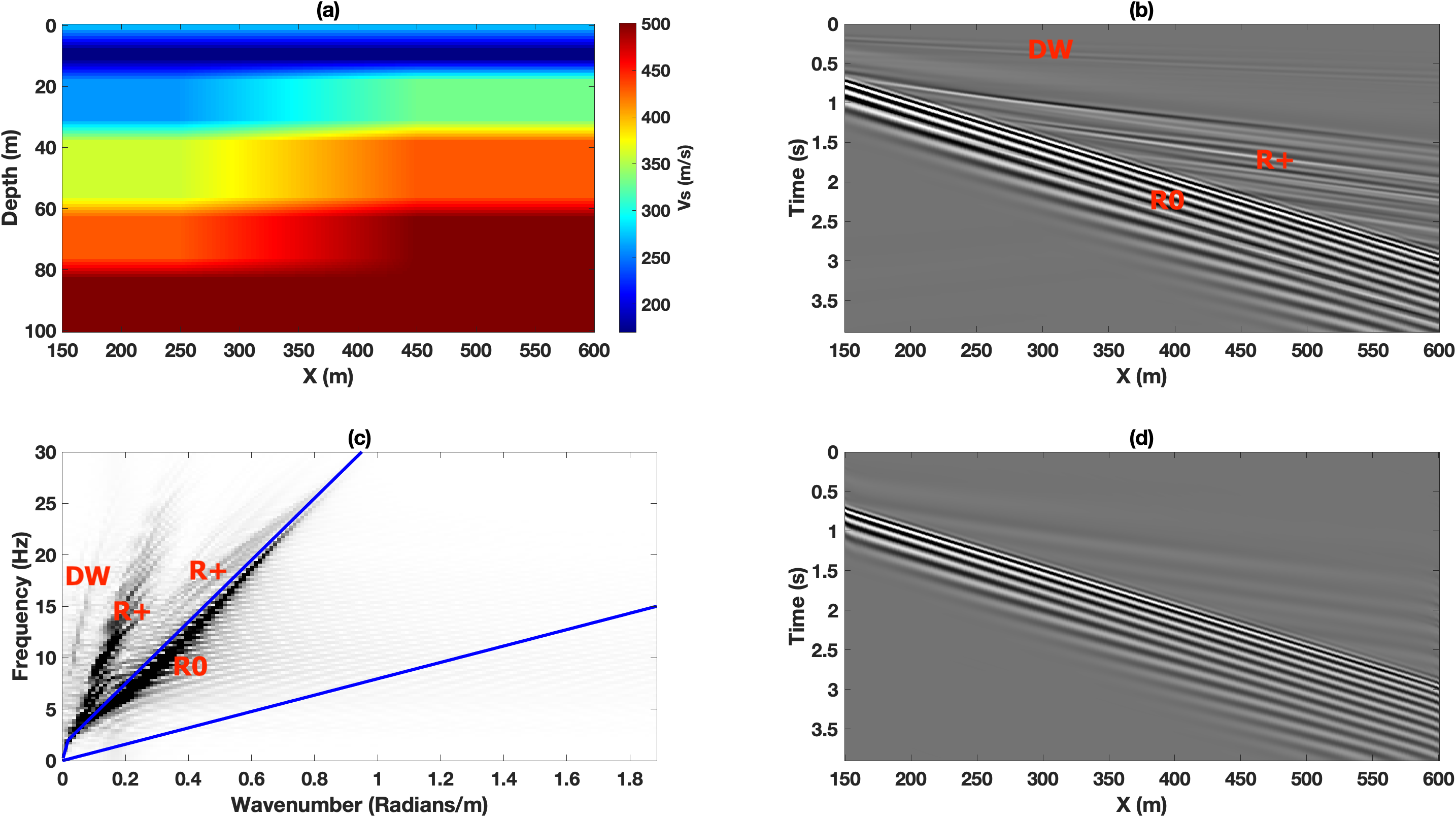}
    \caption{Surface wave spectral analysis of a synthetic active seismic record. (a) The prescribed synthetic model. (b) The horizontal component of the wavefield recorded at the surface for a source located at x = 0 m. (c) $f-k$ spectrum.  fundamental-mode surface waves (R0), higher-modes surface waves (R+), and direct body waves (DW) are indicated in (b-c). (d) An isolated fundamental-mode is retained after an f-k filter has been applied. The blue lines in (c) represent the corner frequencies of the applied velocity filter.}
    \label{fig:Syntheticdata}
\end{figure}

In order to test the proposed workflow, we consider a 2D synthetic model with lateral variation and a low-velocity layer near the surface (Figure~\ref{fig:Syntheticdata}a). The wavefield is modeled using a two-dimensional finite-difference elastic wave equation solver \citep{thorbecke20172d} assuming a free surface at the top of the model and a compressional (p-wave) source type. The source time function is a 12 Hz Ricker wavelet. Since the straight fibers record the radial component of the surface waves, we use the horizontal component of the wavefield recorded at the surface of the synthetic model (Figure~\ref{fig:Syntheticdata}b). 
In this experiment, the recorded surface wave on the horizontal component of the wavefield is the Rayleigh wave. We indicated the fundamental-mode surface wave (R0), higher-modes surface waves (R+), and the direct body wave (DW) on the seismic record in Figure~\ref{fig:Syntheticdata}b. The $f-k$ spectrum is shown in Figure~\ref{fig:Syntheticdata}c. 
The fundamental-mode Rayleigh wave is dominant in Figure~\ref{fig:Syntheticdata}b-d. 

\subsection{Multi Offset Phase Analysis}

\begin{figure}[h]
    \centering
    \includegraphics[width=1\linewidth]{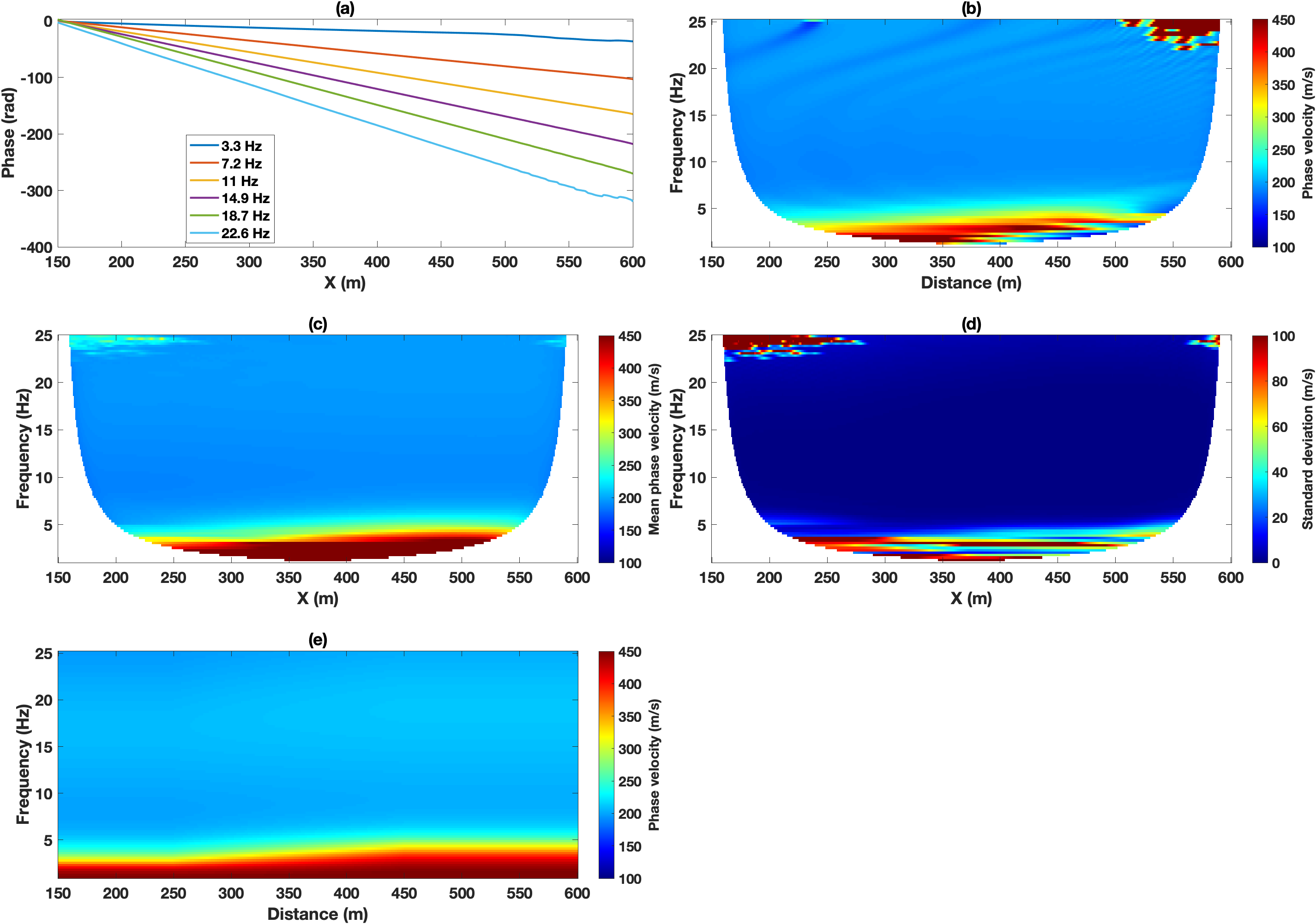}
    \caption{Phase velocity retrieval using MOPA: application to synthetic recordings. (a) Phase versus offset for six discrete frequencies of the filtered shot record presented in Figure~\ref{fig:Syntheticdata}d. (b) Phase velocities $c(f_i,x_j^{\mathrm{(c)}})$ retrieved from one single shot record. (c) Mean local phase velocities retrieved from 32 shots. (d) Standard deviation corresponding to (c). (e) Theoretical (true) laterally varying phase velocities.}
    \label{fig:MOPA_Syntheticdata}
\end{figure}

To obtain a reliable phase versus offset spectrum that is associated with the fundamental-mode only, we isolate the fundamental-mode (Figure~\ref{fig:Syntheticdata}d) by retaining the energy contained by the blue lines in Figure~\ref{fig:Syntheticdata}c and muting the rest of the spectrum ($f-k$ filtering). The filtered data, after computation of the inverse transform, is presented in Figure~\ref{fig:Syntheticdata}d. Subsequently, each trace of the ``cleaned'' shot record is transformed to the frequency domain using a Fast Fourier Transform. For each discrete frequency $f_i$, the phases of the individual traces are unwrapped \citep[e.g.,][]{Weemstra2020}, resulting in the $\phi-x$ spectrum. The (unwrapped) phase versus offset is presented for six different frequencies in Figure~\ref{fig:MOPA_Syntheticdata}a. 

The retrieved laterally varying dispersion curves using the MOPA are depicted in Figure~\ref{fig:MOPA_Syntheticdata}b for the single shot record of Figure~\ref{fig:Syntheticdata}d. Here, we used a spatial window length $W(f_i)$ equal to two wavelengths, where the latter is computed using a reference phase velocity based on the (averaged) phase velocity retrieved through the application of MOPA to the whole shot record. This frequency dependency implies that $W$ decreases with increasing frequency. Linear regression using Equation \eqref{eq:linearphaseoffset} for each $f_i$ and $x_j^{\mathrm{(c)}}$ separately, subsequently results in the frequency-dependent ``local" wavenumbers. These wavenumbers are then transformed into phase velocities using Equation \eqref{eq:phasevelocitywavenumber}, yielding the set of laterally varying dispersion curves depicted in  Figure~\ref{fig:MOPA_Syntheticdata}b.

To improve the quality of the recovered dispersion curves, MOPA is conventionally applied to multiple shots, each located at a different (in-line) position. In this synthetic experiment, we modeled 32 shot records with sources located between 0-150 m and 600-750 m and with a source spacing of 10 m. Additionally, an $f-k$ filter is applied to each (Fourier transformed) shot record to facilitate a reliable phase analysis by isolating the fundamental-mode. Subsequently, laterally varying dispersion curves are estimated through the application of MOPA to each shot record separately. At each position $x_j^{\mathrm{(c)}}$, this results in 32 independently estimated dispersion curves. The average of these 32 sections is presented in Figure~\ref{fig:MOPA_Syntheticdata}c. The associated standard deviation is presented in Figure~\ref{fig:MOPA_Syntheticdata}d, which is a measure of the uncertainty of the recovered $c(f_i,x_j^{\mathrm{(c)}})$. 

Figure~\ref{fig:MOPA_Syntheticdata}e shows the (true) theoretical location-dependent phase velocities as a reference. These are computed by taking the true 1D shear-wave velocity profile at each location and then computing the theoretical dispersion curve using the reduced delta matrix method \citep{buchen1996free,wu2019matlab}. Figure~\ref{fig:MOPA_Syntheticdata}c,e show that
lower frequencies (less than 3 Hz) and also higher frequencies (higher than 20 Hz) are associated with higher uncertainties and deviate from the theoretical dispersion relation. These are due to the low amplitude of the source time function (i.e., a 12 Hz Ricker) at those frequencies. Additionally, residual higher-modes (visible in Figure~\ref{fig:Syntheticdata}d) also affect the quality of the dispersion relation. Finally, the MOPA algorithm does not cover the regions close to the profile ends due to spatial windowing. Since this window length $W$ is larger at lower frequencies, lower frequencies sampled over a shorter spatial interval in Figure~\ref{fig:MOPA_Syntheticdata}b,c.

\subsection{Inversion}
\begin{figure}[h]
    \centering
    \includegraphics[width=1\linewidth]{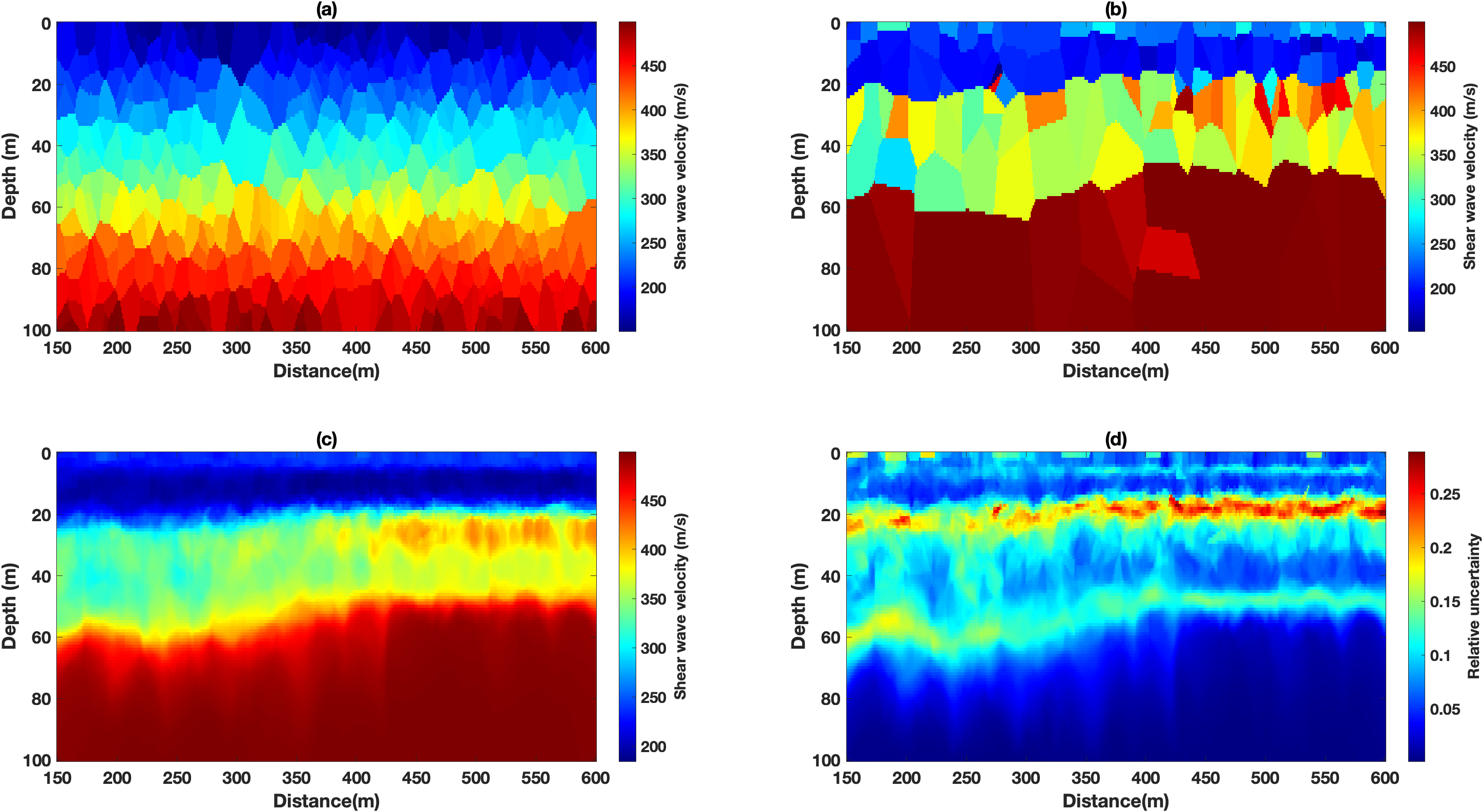}
    \caption{(a) A random initial model of a chain. (b) The last collected sample of the chain. (c) The posterior mean of the collected samples from 20 parallel chains. (d) The posterior standard deviation of the retained samples.}
    \label{fig:InversionSynthetic}
\end{figure}

The recovered local phase velocities are now used in the proposed 2D transdimensional inversion algorithm. We used 20 independent chains to sample the posterior each sampling 700,000 models. The initial model of each chain is generated randomly with a randomly selected number of cells and a randomly chosen location. We only assumed an increasing velocity with depth for the initial model (Figure~\ref{fig:InversionSynthetic}a). The last collected sample of that chain is shown in Figure~\ref{fig:InversionSynthetic}b. We discarded the first 200,000 samples as the burn-in period. Then samples are retained at every 100 iterations to avoid collecting correlated samples. Consequently, a total of 100,000 samples are retained and used for the calculation of the posterior mean (Figure~\ref{fig:InversionSynthetic}c) and the posterior standard deviation (Figure~\ref{fig:InversionSynthetic}d).

Figure~\ref{fig:InversionSynthetic}c shows that the proposed algorithm successfully recovered the true shear wave velocity model of Figure~\ref{fig:Syntheticdata}a near the surface. In deeper parts of the model, the recovered shear wave velocity is a smoother version of the true velocity model. The uncertainty presented in Figure~\ref{fig:InversionSynthetic}d is also meaningful by having higher values in the layer interfaces.

\section{Application to DAS data recorded near Zuidbroek, Groningen}
In this section, we discuss the application of the proposed methodology to a field data set recorded using a straight fiber DAS system. We first introduce the data. Then we recover the local phase velocities followed by a 2D transdimensional inversion. The results are then compared with the shear wave velocity profiles measured at two boreholes along the acquisition line.
\subsection{Data characteristics}
\begin{figure}[h]
    \centering
    \includegraphics[width=1\linewidth]{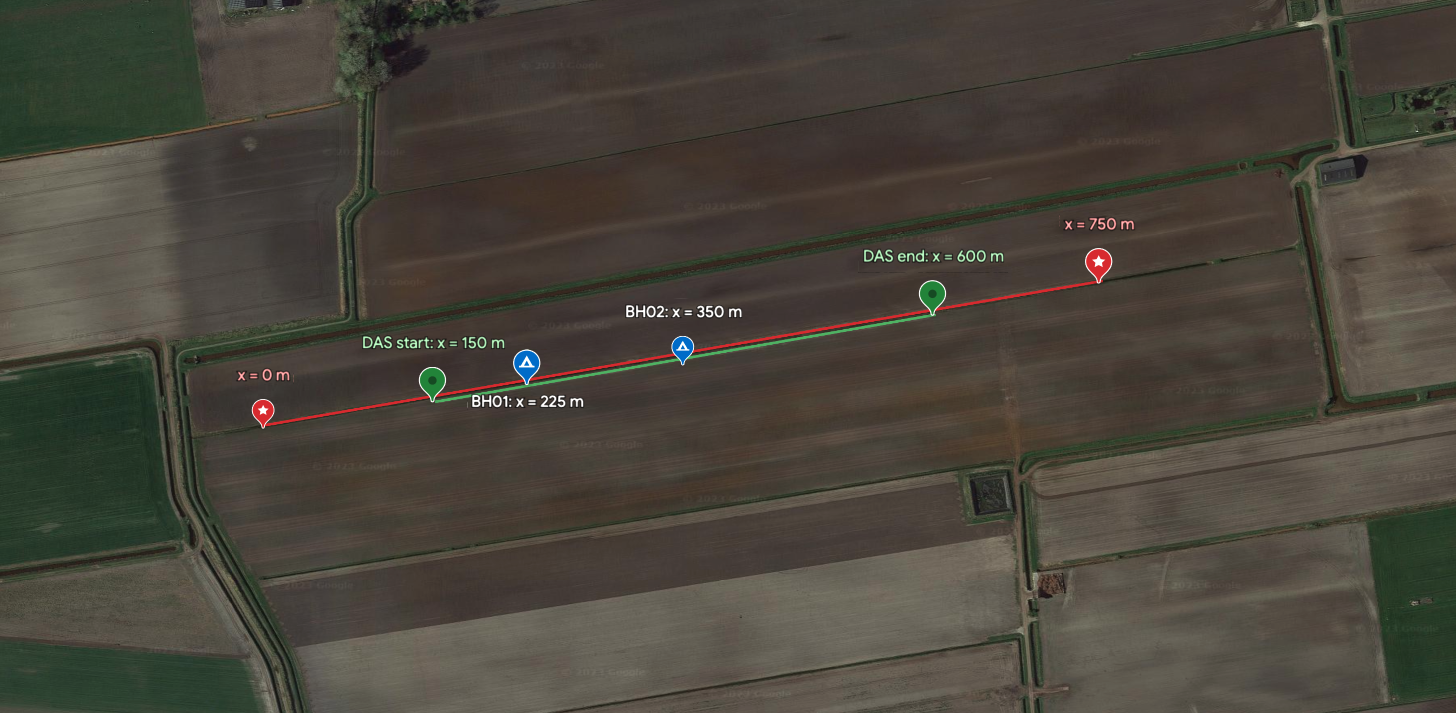}
    \caption{The acquisition setup of the recorded DAS data in the Groningen area. The red line shows the source line with a length of 750 m and a source spacing of 2 m. The green line shows the DAS straight fiber buried 2 m in depth and  with a length of 450 m. The start of the source line is marked as x = 0 and the end of the source line is marked as x = 750 m. The DAS fiber is also located at x = 150 - 600 m. Two boreholes are also drilled to a depth of 80 m at x = 225 m and x = 350 m to measure the shear-wave velocity directly.}
    \label{fig:datasetup}
\end{figure}

The proposed methodology has been applied to data obtained in the area of  Groningen in the north of The Netherlands. These data were obtained with fibers as part of a Distributed Acoustic Sensing (DAS) system. Figure 3 shows the acquisition setup of the DAS system. 
Figure~\ref{fig:datasetup} shows the acquisition setup of the DAS recording system. Several configurations of fibers were used, namely straight and helically wound fibers during the recording \citep{hasani2023experiences}, however, we opted for the straight fiber data as it showed the highest sensitivity to the surface waves.   
The source used is an electrically driven vertical seismic vibrator \citep{Noorlandt2015} shooting 2 shots per position from x=0 to x=750 every 2 m. The straight fiber is buried 2m in depth with a length of 450 m from x = 150-600. More information on the acquisition can be found in \citet{hasani2023experiences}. 

The receiver spacing of the recordings is 1 m with a gauge length of 2 m. We considered 152 off-end shots on the two sides of the recording line for the application of MOPA. In fact, for the purpose of surface wave analysis and inversion, all shots located at x = 0-150 m and x = 600-750 m are considered in this study, a total of 152 shots. 

\begin{figure}[h]
    \centering
    \includegraphics[width=1\linewidth]{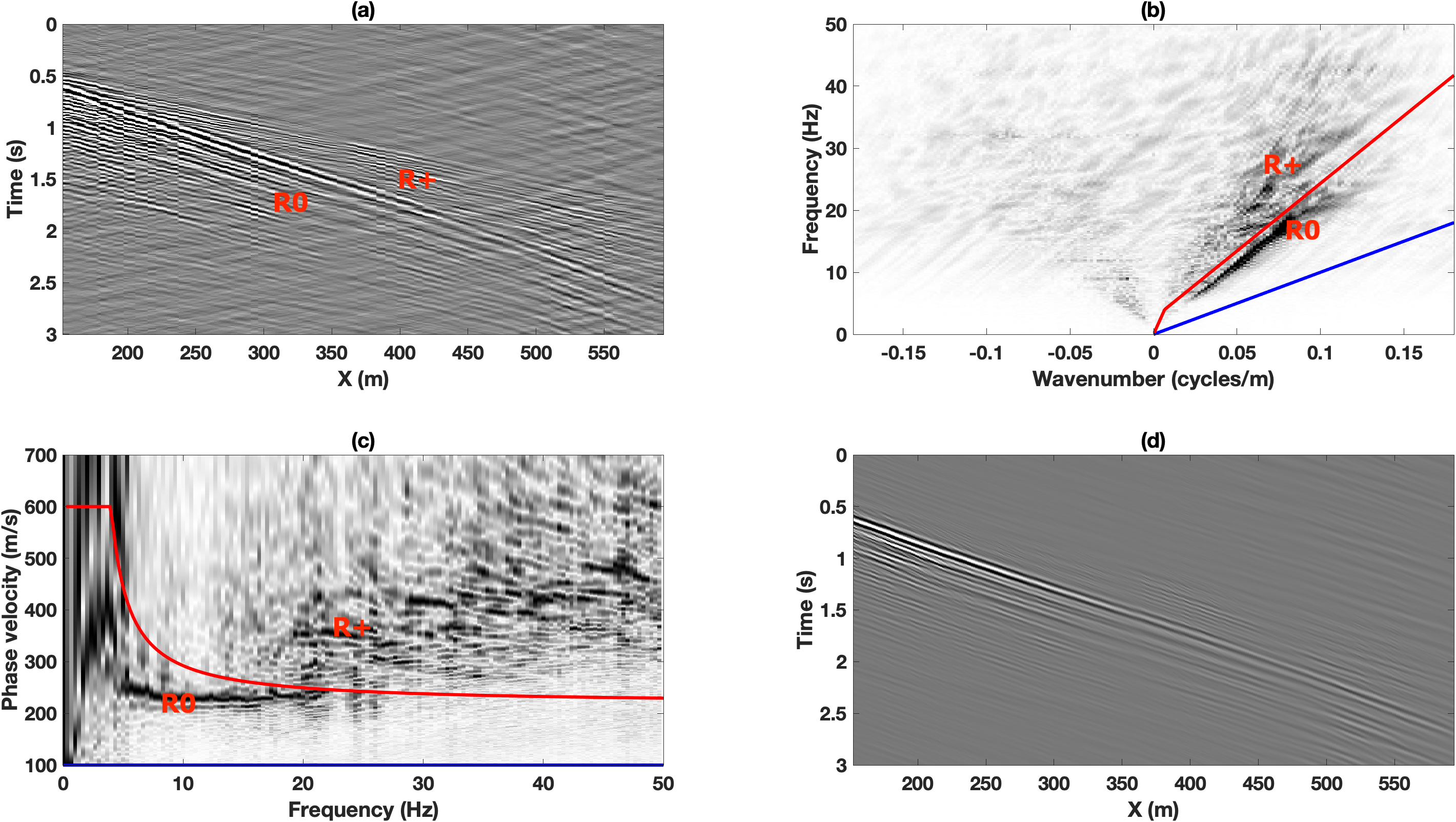}
    \caption{A common shot record (a), its corresponding f-k amplitude spectrum (b), and its corresponding dispersion image (c). The source is located at x = 10 m. The fundamental-mode and higher-modes are indicated by R0 and R+, respectively. The blue line indicates the phase velocity of 100 m/s and the red line indicates a frequency-dependent phase velocity between 220-600 m/s. (d) fundamental-mode surface wave shot records  corresponding to the shots depicted in (a). An f-k filter is applied to the shot records by keeping the f-k spectrum between the blue and red lines. The rest of the spectrum is filtered out.}
    \label{fig:ShotsampleFieldData}
\end{figure}

Figure \ref{fig:ShotsampleFieldData} shows a sample shot gather of the DAS data recorded at a farm in the Groningen area, the Netherlands. The $f-k$ and $c-f$ spectra are also provided for a better understanding of the data. First, a straight fiber records the radial component of the wave field. Since the vertical source is in line with the fiber (i.e., receivers), the recorded surface wave in the radial direction is the Rayleigh wave.
Second, the shot record is dominated by the fundamental-mode Rayleigh wave indicated by \textbf{R0}. The fundamental-mode is easily detectable in the frequency range of 4-20 Hz in both $f-k$ and $c-f$ spectra. Third, higher-modes are also clearly visible in both shot records and their corresponding spectra. We have indicated higher-modes by \textbf{R+} since more than one higher-mode is visible in the $f-k$ and $c-f$ spectra; it is also difficult to separate them. Finally, the colored lines in the $f-k$ spectra represent phase velocities to design a velocity filter for isolating the fundamental-mode. The velocity lines are also depicted on the $c-f$ spectra for the reader's reference. As you can see, the red line separating the fundamental-mode from higher-modes is velocity dependent. The $f-k$ spectrum between the blue and the red line is preserved and the rest of the spectrum is filtered out.  The filtered record is presented in Figure~\ref{fig:ShotsampleFieldData}d representing the fundamental-mode Rayleigh wave. We have applied this velocity filter to all the shots.     

\subsection{Multi Offset Phase Analysis}

\begin{figure}[h]
    \centering
    \includegraphics[width=1\linewidth]{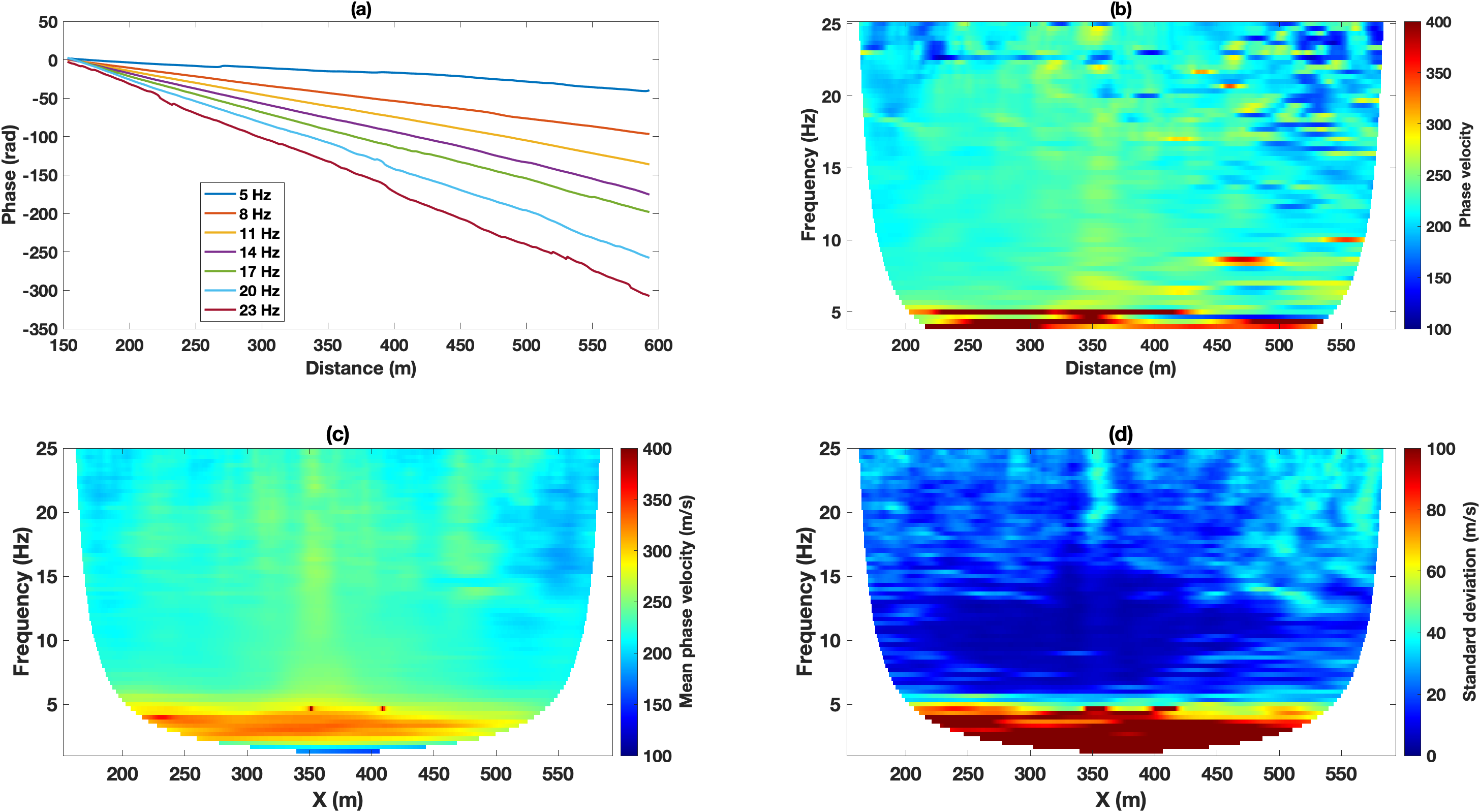}
    \caption{MOPA applied to the field DAS data considering the lateral variation of the subsurface. (a) Unwrapped phase versus offset at seven different frequencies (i.e., different colors) for the filtered seismic record depicted in Figure~\ref{fig:ShotsampleFieldData}d. (b) laterally varying phase velocity retrieved from the single shot record of Figure~\ref{fig:ShotsampleFieldData}d. (c) Mean of the laterally varying dispersion curves retrieved from 152 off-end shots. (d) Standard deviation corresponding to (c).}
    \label{fig:MOPA_fielddata}
\end{figure}

After isolating the fundamental-mode (Figure~\ref{fig:ShotsampleFieldData}d), we applied the MOPA algorithm to all 152 off-end shot records of the available DAS data. Figure~\ref{fig:MOPA_fielddata} shows the results of the MOPA method applied to the field DAS data to retrieve the dispersion curves. The unwrapped phase versus offsets for 7 frequency components of a single shot record (Figure~\ref{fig:ShotsampleFieldData}b) are depicted in Figure~\ref{fig:MOPA_fielddata}a. The retrieved laterally varying dispersion curves are presented in Figure~\ref{fig:MOPA_fielddata}b, which are somehow noisy with rather sharp changes. A more smooth and more reliable dispersion curve can be derived by repeating the process for multiple shot records. 
Figure~\ref{fig:MOPA_fielddata}c presents the average laterally varying dispersion curves derived from 152 shot records. The uncertainty (i.e., standard deviation) is also depicted in Figure~\ref{fig:MOPA_fielddata}d.

\subsection{Prior information}

\begin{figure}[h]
    \centering
    \begin{subfigure}[tb]{0.49\textwidth}
         \centering
         \caption{BH01}
         \includegraphics[width=1\textwidth]{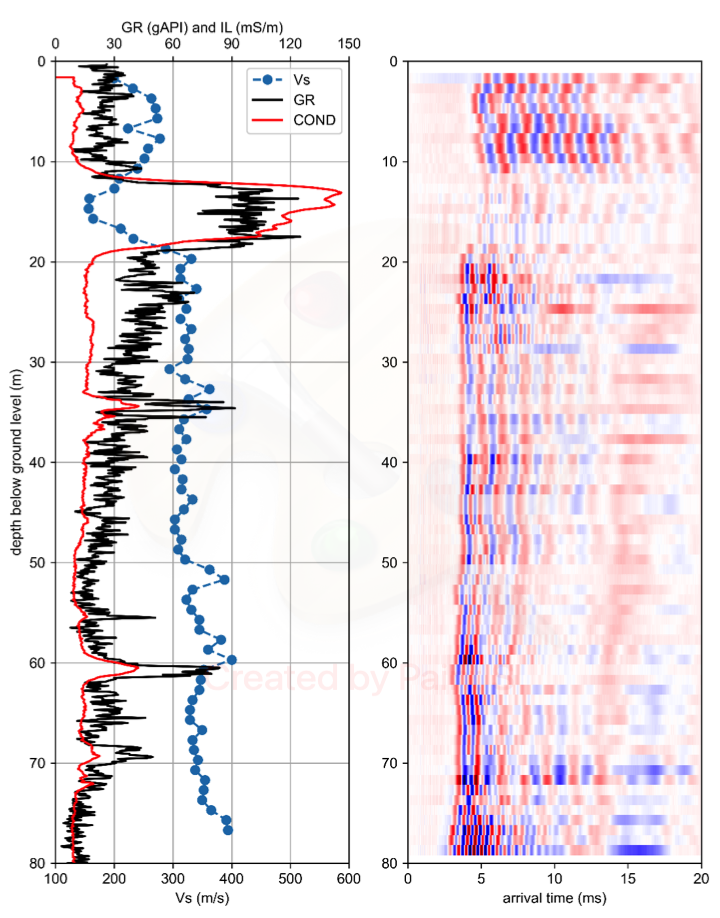}
         \label{fig:BH01}
     \end{subfigure}
     \begin{subfigure}[tb]{0.49\textwidth}
         \centering
         \caption{BH02}
         \includegraphics[width=1\textwidth]{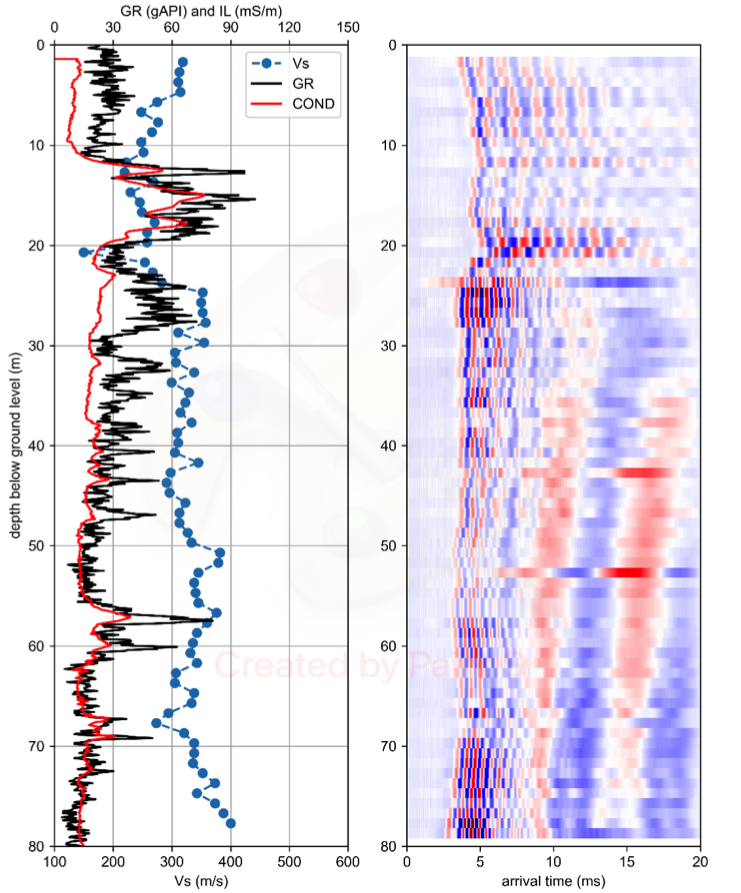}
         \label{fig:BH02}
     \end{subfigure}
    \caption{The measured Induction-Log readings (IL), Gamma Ray (GR), and shear-wave velocity profile ($V_s$) together with the far receiver s-wave seismogram of (a) first borehole (BH01) at x = 220 m and (b) second borehole (BH02) at x = 350 m of the recording line.}
    \label{fig:boreholelogs}
\end{figure}

Prior information is the known information from the area of study based on the possible previous studies or the direct measurement of the subsurface properties. In this study, two open boreholes were drilled down to a depth of 80 m, where the shear-wave velocity and p-wave velocity were measured directly using a PS suspension logging tool. In addition to the seismic data, the bulk electrical conductivity and the gamma radiation were also measured. The measured logs are presented in Figure~\ref{fig:boreholelogs}. The main feature observable from the logs is a clear reduction of shear-wave velocity between 10 and 20 m, especially in Figure~\ref{fig:BH01}. This is due to the higher clay content supported by the conductivity and gamma-ray values having higher values than above and below that layer. In addition, the relatively low electrical conductivity values indicate fresh water and the relatively low gamma ray values indicate a low level of clay content. Consequently, the top 10 m of the subsurface is predominantly sand with fresh water.

\begin{figure}[h]
    \centering
    \includegraphics[width=1\linewidth]{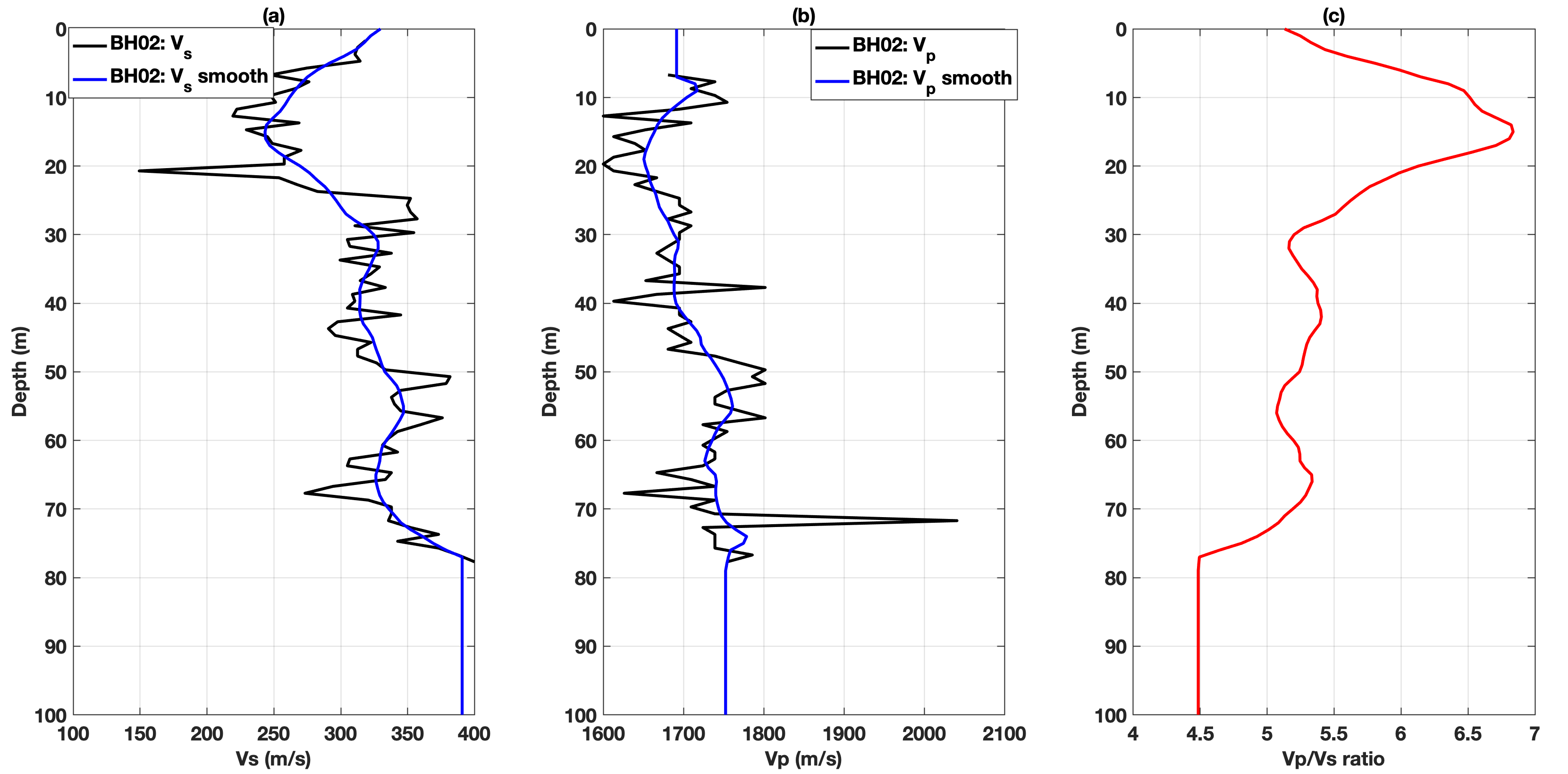}
    \caption{shear-wave velocity profile (a) and p-wave velocity profile (b) measured at the second borehole (BH02). The blue curves are the smooth version of the original velocity profiles using a moving average filter. (c) The $\dfrac{V_p}{V_s}$ ratio is computed based on the smooth blue curves.}
    \label{fig:vpvsratio}
\end{figure}

This logging information can be used to construct a prior probability distribution for the shear-wave velocity, which is the subject of another study. In this study, we assumed an uninformative uniform prior distribution for the shear-wave velocity with the prior bounds of 50-600 m/s, and we used these logging data to validate our inversion results. 
Additionally, to compute the surface wave dispersion curves theoretically, we need the shear-wave velocity ($V_s$), p-wave velocity($V_p$), and density ($\rho$). However, the shear-wave velocity is mainly controlling the theoretical dispersion curves, and the effect of $V_p$ and $\rho$ is ignored in the literature \citet{wathelet2008improved}. 
We used the logging data to get an estimate of the $\dfrac{V_p}{V_s}$ ratio to be used in the theoretical calculation of the dispersion curves. To that end, we first smoothed the shear-wave velocity profile (Figure~\ref{fig:vpvsratio}a) and the compressional-wave velocity profile (Figure~\ref{fig:vpvsratio}b) using a moving average profile. Then, a depth-dependent $\dfrac{V_p}{V_s}$ ratio is computed based on the smoothed blue curves presented in Figure~\ref{fig:vpvsratio}c. During the inversion process, we used a $\dfrac{V_p}{V_s}$ ratio of 5 to relate the shear-wave velocity and the compressional-wave velocity while computing the dispersion curves.

s\subsection{Transdimensional inversion}

The retrieved laterally varying phase velocities (Figure~\ref{fig:MOPA_fielddata}e) are now used in the 2D transdimensional algorithm. We used 20 independent McMC chains running in parallel each sampling 700,000 models. The initial model at each chain is selected randomly with velocity increasing with depth. The first 200,000 samples of each chain are discarded as the burn-in phase. To avoid correlated samples, the samples are retained after every 100 iterations of the McMC. Consequently, a total of 100,000 retained samples are used as the representation of the posterior distribution to compute the posterior mean (i.e., posterior expectation) and standard deviation (i.e., posterior uncertainty). 

\begin{figure}[h]
    \centering
    \includegraphics[width=1\linewidth]{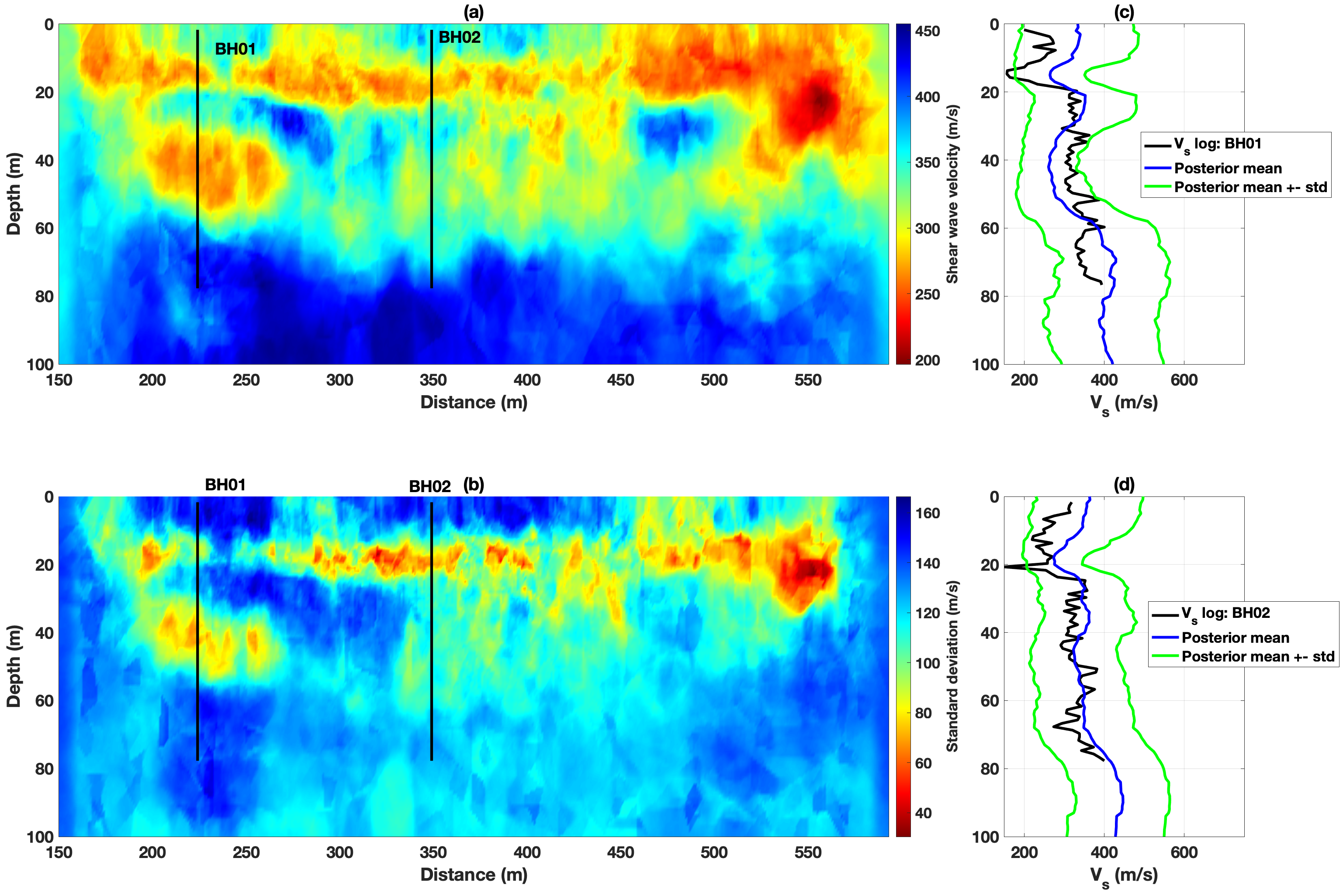}
    \caption{2D transdimensional inversion results. (a) Recovered shear-wave velocity 2D section (i.e., posterior mean), (b) recovered model uncertainty (i.e., posterior standard deviation) and comparison with the shear-wave velocity profile available from the first borehole (c) and the second borehole (d).}
    \label{fig:TransD_InvFielddata}
\end{figure}

Figure~\ref{fig:TransD_InvFielddata} presents the recovered shear-wave velocity profile using the 2D transdimensional method. The shear-wave velocity profiles from the well logs are also compared with the transdimensional inverted result. As one can see, the inverted shear-wave velocity is in good agreement with the well log data at the location of the second borehole (i.e., x = 350 m). The inverted result follows the well log data and the main anomalies are resolved. At the location of the first borehole (i.e., x = 220 m), the inverted shear-wave velocity deviates a bit more from the well log data while predicting the low velocity anomaly between the depth of 10-20 m. This deviation at the location of the first borehole is mainly due to the lower quality of the MOPA derived dispersion curves. As one can see in Figure~\ref{fig:MOPA_fielddata}e-f, the retrieved phase velocity at the sides of the model has fewer frequency components and is associated with higher uncertainties.

\section{ Discussion}

There are several points of discussion. First, the agreement between the transdimensional derived shear-wave velocity and the well logs are promising, especially at the location of the second borehole close to the middle of the seismic line. However, at the location of the first borehole where the retrieved phase velocity is missing lower frequencies and associated with higher uncertainties, the results deviate from the more reliable and locally measured log model. This is due to the MOPA algorithm we used in this study. The quality of the dispersion curves at these locations might be improved by the tomography-derived phase velocity retrieval algorithm \citep{barone2021tackling}. 

Second, data uncertainty ($\sigma$ in Equation~\ref{eq:Likelihood}) plays a crucial role in the convergence of a Bayesian algorithm \citep{bodin2012transdimensionalunknownnoise,rahimi2021potential}. As discussed earlier, the MOPA algorithm we used in this study provides us with an estimate of the data uncertainty (Figure \ref{fig:MOPA_Syntheticdata}f). In this study, we used this MOPA-derived uncertainty during the McMC sampling of the posterior. However, modeling and other processing errors are also affecting Bayesian inference. Consequently, we suggest further studies to take into account these errors to improve the convergence of each sampling chain and to enhance the quality of the final inversion result.

Third, the proposed 2D transdimensional algorithm is comparable with the independent 1D inversion of the dispersion curves in terms of computational time 
This is because in the proposed 2D transdimensional algorithm, at each step of the McMC, we perturb a small portion of the model space. Then, we compute the forward function (i.e., the primary source of the computational demand) only at the updated part of the model space. This reduces the computational time significantly. The added values of the proposed 2D transdimensional scheme are the enhanced lateral correlation and the reduced solution nonuniqueness.

Finally, we assumed a uniform prior whereas we have two boreholes across the line measuring the shear-wave velocity directly. Using this kind of available prior data can improve the inversion result significantly.


\section{Conclusion}

We investigated the potential of straight fiber DAS data in combination with a 2D transdimensional inversion algorithm to recover a 2D reliable shear-wave velocity section. We successfully recovered the laterally varying phase velocities using Multi Offset Phase Analysis. Then, we adopted the 2D transdimensional algorithm to invert all the available laterally varying dispersion curves simultaneously. The recovered shear-wave velocity is smooth with imaging of the lateral variation of the near-surface. The inverted shear-wave velocity section indicates a low-velocity anomaly between 10-20 m depth which is in agreement with the shear-wave velocity logs. The low velocity is due to the high content of clay supported by gamma-ray measurements. In general, the recovered shear-wave velocity section matches the log at the second borehole at 10-65 m depth. Therefore, we were able to retrieve the lateral variability of the subsurface using Rayleigh waves, with a good match with the S-wave logs in two boreholes.

\section*{Acknowledgements}
This research has received funding from the European Research Council (ERC) under the European Union’s Horizon 2020 research and innovation program (grant no. 742703).
\clearpage 
\newpage
\bibliography{references.bib}

\newpage

\end{document}